# Superconducting state of quasiparticles with spin dependent mass and their distinguishability for Cooper-pair state

J. Spałek[1*], M. M. Maśka[2], M. Mierzejewski[2], and J. Kaczmarczyk[1]

[1]Marian Smoluchowski Institute of Physics, Jagiellonian University, Reymonta 4, 30-059 Kraków, Poland

[2]Institute of Physics, University of Silesia, Uniwersytecka 4, 40-007 Katowice, Poland

[*]e-mail: ufspalek@if.uj.edu.pl

**Spin dependence of quasiparticle mass has been observed recently in CeCoIn$_5$ [1] and other systems[2,3]. It emerges from strong electronic correlations in a magnetically polarized state and was predicted earlier[4,5,6]. Additionally, the Fulde-Ferrell-Larkin-Ovchinnikov (FFLO) phase[7,8,9] has also been discovered in CeCoIn$_5$ [10,11,12] and therefore, the question arises as to what extent these two basic phenomena are interconnected, as it appears in theory (see METHODS). Here we show that the appearance of the spin-split masses essentially extends the regime of temperature and applied magnetic field, in which FFLO state is stable, and thus, it is claimed to be very important for the phase detectability. Furthermore, in the situation when the value of the spin quantum number σ=±1 differentiates masses of the particles, the fundamental question is to what extent the two mutually bound particles are indistinguishable quantum mechanically? By considering first the Cooper-pair state we show explicitly that the antisymmetry of the spin-pair wave function in the ground state may be broken when the magnetic field is applied.**

The newly discovered heavy-fermion superconductors are termed as unconventional because of the basic symmetries such as spatial inversion[13] or time reversal[14,15] are broken in some cases. Here we examine one of the basic new features of those systems, namely, the spin dependence of quasiparticle mass, which appears in the applied magnetic field H≠0. The basic observation is that by switching on the applied field one may transform the system of quantum-mechanically indistinguishable quasiparticles into their distinguishable correspondents. This circumstance, in turn, produces not only quantitative changes of normal-state properties, but also leads to a qualitative modification of the single-Cooper-pair state composed of such distinguishable particles, as discussed explicitly below.

Generally, the spin dependent masses of carriers should appear in narrow band systems in the situation with a net magnetic moment per site, $<m> = <n_\uparrow - n_\downarrow>$, because of the presence of strong electronic correlations driven by the short-range Coulomb repulsion, which has magnitude comparable (or even larger) than the single-particle (band) energies near the Fermi level. Explicitly, the magnitude of interaction is characterized by the so-called Hubbard term $U\Sigma_i n_\uparrow n_\downarrow$ (see METHODS). The mass renormalization in the $U\rightarrow\infty$ limit obtained nonperturbatively[5,6,7,16] amounts to $m_\sigma^*/m_0=(1-n_\sigma)/(1-n)$, where $m_0$ is the bare (band) mass, $n_\sigma=<n_{i\sigma}>$ is the number of carriers (per correlated state) with spin $\sigma$, and $n=n_\uparrow+n_\downarrow$ is the total number of electrons per state, the so-called band filling. The effect is particularly strong near the half-filling of the narrow band, i.e. $n\rightarrow 1$, when $n=1-\delta$ with $\delta<<1$, and the denominator becomes almost divergent. Strictly speaking, the prototypical situation arises when a very narrow band of heavy quasiparticles ($m^*/m_0\sim 10^2$) results from hybridization of originally atomic $4f^1$ states of $Ce^{3+}$ ions (one carrier per ion) with uncorrelated (itinerant) 5d-6s electrons. In that situation, the band filling n is played by the filling $n_f$ of the 4f level in the compound and $\delta\equiv 1-n_f$ describes slight (<5%) deviation in the intermetallic compound from the atomic $Ce^{3+}$ valency ($Ce^{3+}\rightarrow Ce^{(3+\delta)+}$). Starting from the paramagnetic case, when $m_\uparrow^*=m_\downarrow^*\equiv m_{av}=(1-n_f/2)/(1-n_f)$ and switching on the field H, the mass difference $\Delta m^*/m_0=<m>/(1-n_f)$ is predicted to be linear in magnetization[17]. One has to underline that the mass enhancement $m_\sigma^*/m_0$ is momentum-independent and is the additional factor to the band-state Zeeman splitting (when H≠0). This factor will also lead to an essential modification of the original FFLO state[7,8], as well as modify the Cooper-pair state[18] discussed first.

Formulation of the pair problem with masses $m_\uparrow^* \equiv m_1$ and $m_\downarrow^* \equiv m_2$ is similar to that for the standard case, except here we consider also the pair states with the centre-of-mass (COM) momentum $\mathbf{Q} \equiv \mathbf{k}_1 + \mathbf{k}_2 \neq 0$, as well as have to introduce a relative momentum in a nontrivial manner, namely[19] $\mathbf{k} \equiv (\mathbf{k}_1 m_2 - \mathbf{k}_2 m_1)/(m_1 + m_2)$. We consider the pair wave function separated into the spatial $\Phi(\mathbf{r}_1, \mathbf{r}_2)$ and spin $\chi_{\sigma_1,\sigma_2}(1,2)$ parts

$$\Psi(\mathbf{r}_1, \mathbf{r}_2, \sigma_1, \sigma_2) = \Phi(\mathbf{r}_1, \mathbf{r}_2) \chi_{\sigma_1,\sigma_2}(1,2) \qquad (1)$$

We select the spin function in either the singlet form

$$\chi_{\sigma_1,\sigma_2}(1,2) = \frac{1}{\sqrt{2}}\left[\chi_\uparrow(1)\chi_\downarrow(2) - \chi_\downarrow(1)\chi_\uparrow(2)\right] \qquad (2)$$

reflecting the particle *indistinguishability*, or in the form

$$\chi_{\sigma_1,\sigma_2}(1,2) = \chi_\uparrow(1)\chi_\downarrow(2) \quad \text{or} \quad \chi_\uparrow(2)\chi_\downarrow(1) \qquad (3)$$

reflecting the particle *distinguishability* by their spin-direction-dependent masses in the corresponding σ=↑,↓ states. The spin distinguishability originates in the Hamiltonian for two particles, in which the external characteristic – the effective mass – depends on spin. In Fig. 1 we have shown the difference between the singlet state (top) and the two-particle state of distinguishable particles; they transform differently under the spin transposition and the latter state has no definite symmetry in this respect, as marked. It turns out that the pair at rest (**Q**=0) has the wave function given by Eq.(2) antisymmetric with respect to the transposition, whereas that with **Q**≠0 has not.

The resulting single-Cooper-pair properties are summarized in Fig. 2. In part (**a**) we plot the pair binding energy as a function of applied field: The green and blue solid lines are drawn respectively for the case with the antisymmetric function for **Q**=0 and given by Eq.(3) for **Q**≠0, respectively. The dashed lines represent the corresponding solutions for $m_\sigma^* = m_{av}$ (with spin independent masses) and respectively with the total wave function antisymmetric (green line) and without that symmetry (blue). One sees clearly that the energetically stable solution violates not only the spin transposition symmetry, but also the antisymmetric character of the total wave function (1), i.e. with respect to the transposition of complete coordinates, $(\mathbf{r}_1,\sigma_1) \leftrightarrow (\mathbf{r}_2,\sigma_2)$. This is, in fact, relatively easy to understand as their $\sigma$-dependent masses are the extra characteristics of the effective (quasiparticle) approach. In Fig. 2b we show the value of momentum $|\mathbf{Q}|$ for a stable solution as a function of H and mark explicitly two critical fields: the field, at which the solution with $Q \cong |k_{F\uparrow} - k_{F\downarrow}|$ appears (the blue point in Fig. 2a), as well as the applied field destroying the bound pair state with opposite spins (red balls in Figs. 2a and 2b). For lower fields, the solution with $|\mathbf{Q}_{interm}| \neq 0$ appears (the $\mathbf{Q}=0$ solution is stable only at H=0). In Figs. 2c and d we characterize the solutions with $\mathbf{Q} \neq 0$ in a fixed field; note also the rapid decrease of the energy for the solution with $m^* = m_{av}$ and the antisymmetric wave function for $\mathbf{Q}=0$, cf. the dotted line in Fig. 2d. Finally, we display in Fig. 2e the asymmetry factor i.e. the ratio of the integrated antisymmetric to symmetric parts squared of the spatial wave function, as a function of the applied field. The asymmetry factor quantifies the admixture of the spatially antisymmetric ("improper") part to the symmetric part. Clearly, none of the three factors in expression (4) has a definite transposition symmetry when $m_1 \neq m_2$, as does not the total wave function. So, the considered Cooper pair state for $H \neq 0$ is an example of a quantum state of *distinguishable* particles.

We consider now the condensed state of pairs, both the BCS-type state and that corresponding to the Cooper-pair state with $\mathbf{Q} \neq 0$ (the state is called the FFLO state and will reflect the mismatch $\Delta k_F \neq 0$ for $H \neq 0$). The Cooper state of distinguishable particles for $H \neq 0$ does not prevent us from constructing the condensed BCS-like state of identical pairs and composed of heavy quasiparticles with a well defined momentum and the spin dependent masses. The pairing part is of real-space character and appears also naturally in the strong-correlation limit (see METHODS). Having in mind different symmetries of the superconducting gap, we assume that $\Delta_\mathbf{k} = \Delta_\mathbf{Q} \eta(\mathbf{k})$, where $\eta(\mathbf{k})$ depends on the choice of the gap symmetry and is superimposed on the $\mathbf{Q}$ dependence the gap amplitude[20] $\Delta_Q$. In effect, the two possible branches of quasiparticle energies acquire the form

$$E_{k,Q,\pm} = g\mu_B H + \frac{1}{2}\left(\epsilon_{k\uparrow} - \epsilon_{(k+Q)\downarrow}\right) \pm \frac{1}{2}\sqrt{\left[\epsilon_{k\uparrow} + \epsilon_{(k+Q)\downarrow} - 2\mu\right]^2 + 4\left|\Delta_Q \cdot \eta_k\right|^2}, \quad (4)$$

where μ is the system chemical potential and $\epsilon_k$ is the dispersion relation for bare electrons. The factor $\eta(\mathbf{k})$ is here taken in the form of the d-wave, $\eta(\mathbf{k})=\cos(k_x)-\cos(k_y)$, as observed[12,21]. Explicitly, in the two-dimensional case, which is appropriate for CeCoIn$_5$ with the field oriented along the tetragonal (c) axis we have the quasiparticle energy in the form

$$\epsilon_{k\sigma} = \left\{-2t\left[\cos(k_x) + \cos(k_y)\right] + 4t'\left[\cos(k_x)\cdot\cos(k_y)\right]\right\}\frac{1-n_f}{1-n_{f\sigma}}, \quad (5)$$

where t and t' are the hopping matrix elements between the first and the second neighbors, respectively and the last factor is the spin-dependent renormalization factor. Here we assume that the valency $n_f$ does not depend on the magnetic field, as the corresponding metamagnetic transition in that system is well above the second critical field. The energy is minimized also with respect to **Q** for each H and T. Therefore, we do not assume an explicit form of $\Delta_Q$. Also, to determine explicitly the chemical potential μ we assume that the number of particles per site is $n_f$=0.97 and the ratio of t'/t = 0.5. In Fig. 3 we present the normal state spin-split masses as a function of applied field close to the half-filling. Taking into account that the value of linear specific coefficient is γ=200~mJ/K$^2$mol we can estimate the value of t~20 K. The regime of physical fields H<20 T is limited to $g\mu_B H/t \leq 0.1$, where the mass splitting is already essential.

To characterize superconducting phase we have plotted the phase diagram on the temperature-applied field plane in Fig. 4 ab in the situation with the spin-dependent ($m^*_\sigma \neq m^*_{\bar\sigma}$) and -independent ($m^*_\sigma = m_{av}$) masses, respectively. The BCS-like state is robust in lower fields, whereas the FFLO state is stable in much wider field range if the masses are spin-direction dependent. The last feature may be regarded as one of the reasons for the FFLO observability in CeCoIn$_5$. Additionally, as illustrated in Fig. 4c, the BCS→FFLO phase transition is discontinuous while the FFLO→normal state transition is of second order. The discontinuous nature of the **Q**≠0 appearance is shown in Fig. 4d. One should also mention that as |**Q**| is large (|**Q**|~$\pi/a$, a - lattice parameter), our analysis of the FFLO on a lattice is more appropriate than that based on the continuous Ginzburg-Landau model.

To illustrate the nature of the BCS-FFLO transition, we have drawn in Fig. 5 the exemplary profiles of the free energy which depends explicitly on **Q**=($Q_x,Q_y$) in low (Fig. 5a) and high fields (Fig. 5b). The transition between the two states is indeed first order, as the two states always exist and correspond to the two separate local minima located respectively at **Q**=0 and |**Q**|~$\pi/a$, as in the single-pair case. Finally, we should underline that the border lines have been determined in the strong Pauli limiting case[9,21] and its applicability has been checked out explicitly.

We conclude with some remarks about the effects due to the spin-splitting of the quasiparticle mass. First, the particles forming the Cooper pair transform from *indistinguishable* to *distinguishable* when the magnetic field is applied. Such a situation is similar to that appearing[22] in color superconductivity in QCD. This effect may be tested in the pair tunneling experiments and/or by detecting anomalies in the Andreev reflection. Second, there should be an anomalous T dependence of the penetration depth. These effects should prove that even the BCS-like state is unconventional for heavy-fermion and other correlated electron systems.

**Methods**

Our approach bases on the concept of spin dependent quasiparticle mass obtained for both Hubbard (HM) and periodic Anderson (PAM) models[4,5,6] in the saddle-point (or Gutzwiller) approximations in a magnetically polarized state. Those quasiparticles are subsequently subjected to a local (real-space) pairing, which represents an indispensable part, in the limit of large but finite U, as demonstrated next. Namely, the two features can be related directly within the Anderson-lattice model in the so-called Kondo-lattice limit[23,24,25]. Explicitly, in the large-U limit, PAM can be canonically transformed[23] to the following effective Hamiltonian projected onto a subspace without double a-occupancies and with real-space pairing.

$$\tilde{H} = \sum_{mn} t_{mn} c^+_{m\sigma} c_{n\sigma} + \epsilon_f \sum_{i\sigma} N_{i\sigma}(1-N_{i\bar\sigma}) + \sum_{im} V_{im}(1-N_{i\bar\sigma})(a^+_{i\sigma}c_{m\sigma} + c^+_{m\sigma}a_{i\sigma}) - \sum_{imn} \frac{2V_{mi}V_{in}}{U+\epsilon_f} b^+_{im} b_{in} \quad (M1)$$

The first term represents originally conduction (c) electrons (with the hopping $t_{mn}$) which are intermixed with the atomic (a) electrons located at energy $\epsilon_f$, $N_{i\sigma} \equiv a^+_{i\sigma}a_{i\sigma}$. The two subsystems are hybridized with the amplitude $V_{im}$. The essential feature of the approach is that only a part of hybridization (involving highly excited states) is transformed out into the Kondo-like local coupling, which is expressed in terms of real space pairing. The pairing operator has the form

$$b^+_{im} \equiv (b_{im})^\dagger = \frac{1}{\sqrt{2}} \left[ a^+_{i\uparrow}(1-N_{i\downarrow})c^+_{m\downarrow} - a^+_{i\downarrow}(1-N_{i\uparrow})c^+_{m\uparrow} \right] \quad (M2)$$

Note that the pairing disappears if $U=\infty$ is taken literally. This effective model, when brought to the hybridized basis, can be solved in the slave-boson saddle-point approximation[24] or in Gutzwiller[22] approximation, each combined with the BCS-type of decoupling for the paired part. In effect, the projected hybridization part (the third term) is represented by the spin-dependent matrix element $V_{im} \to \tilde{V}_{im} = V_{im} q_\sigma^{1/2}$, and similarly for the separable pairing potential of a more elaborate character, explicit form of which is not relevant here[24,25]. In the *Kondo-lattice limit* and $n_f<1$, the Hamiltonian (M1) can be brought up to the model of pairing in a single very narrow band with the BCS-type of Hamiltonian (up to a constant) of the form

$$H_{BCS} = \left(\frac{V}{\Delta_a}\right)^2 \sum_{k\sigma} q_\sigma \epsilon_k \Psi^\dagger_{k\sigma}\Psi_{k\sigma} - \frac{4V^4}{\Delta_a^2(U+\epsilon_f)}(q_\sigma q_{\bar\sigma})^{1/2} \frac{1}{N}\sum_{kk'}\gamma_k\gamma_{k'}\Psi^\dagger_{k\uparrow}\Psi^\dagger_{-k\downarrow}\Psi_{-k'\downarrow}\Psi_{k'\uparrow}, \quad (M3)$$

where $\Delta_a$ is the distance between the bare f-level position ($\epsilon_f$) and the Fermi energy, V is the amplitude of (interatomic) hybridization, and $\gamma_k$ is complicated function of the quasimomentum (here, for simplicity taken of either extended s- or d-wave forms). $\Psi^\dagger_{k\sigma}$ represents the creation operator of a hybridized state; to a good accuracy $\Psi_{k\sigma} \cong a_{k\sigma}$.

Two important remarks have to be made at this point. First, the original atomic electrons acquire band properties by a three-step process: hopping a→c from atomic f state into conduction band, followed by a propagation in the conduction band, and a subsequent deexcitation c→a. Hence the effective bandwidth is proportional to $(V^2/\Delta)t_{mn}$ and the factor $q_\sigma$ is due to strong correlations in the U=∞ limit (the propagation of f electron takes place only if the final site is empty, i.e. when $n_f<1$). Second, the mechanism of pairing has the second-order correction in V/U and analogous form to that appearing in t-J model, as the formal derivation is quite similar to our original derivation of the former model (for a recent didactical comparison of the two models see[26]). However, unlike in t-J model, here the strongly correlated state can be regarded as a Fermi-liquid state, albeit unconventional, at least in some cases.

The system of resulting integral equations determining the characteristics of the paired state: $\Delta_\mathbf{Q}$, $\mu$, $n_\sigma$, and $\mathbf{Q}$ have been solved numerically by integrating over the two-dimensional Brillouin zone ($k_x$, $k_y$). The results have been checked out additionally by using an independent procedure involving a numerical integration over determined earlier density of states. Such a checkout was required, since the free-energy differences between the BCS and the FFLO phases were small (cf. Fig. 5) and reaching a numerical consistency quite involved.

**Acknowledgement**

The work was supported by Ministry of Science and Higher Education through Grants Nos. 1 P03B 001 29 and 1 P03B 071 30. Also, the work was carried out under auspices of the European Network "*Emergent Behaviour of Correlated Matter*" (COST P-16). J. S. acknowledges also the senior fellowship of the Foundation for Polish Science (FNP).

**Competing financial interest**

The authors declare that they have no competing financial interest.

Figure Legends

Fig. 1.

Schematic representation of spin states of Cooper pair. **a**, For a Cooper pair composed of particles with the same masses (stable when at rest), the spin part of the wave function is antisymmetric with respect to the spins exchange (particles are indistinguishable). **b**, For the moving pair ($\mathbf{Q}\neq 0$) i.e. for the case with the spin dependent masses of quasiparticles, there is no such transposition symmetry (the particles are distinguishable). The field presence thus transforms the indistinguishable quasiparticles into distinguishable objects, as discussed in main text.

Fig. 2.

Characteristics of a single Cooper pair in gas and in applied magnetic field. **a,** Pair binding energy vs. H in several situations. The blue solid point reflects the (lower) critical field for the appearance of the state with $Q\cong|\Delta k_F|$ when $m_\uparrow^* \neq m_\downarrow^*$. The red solid points located on the x-axis denote the (upper) critical field for the disappearance of the bound state. The solid lines reflect the bound state energy when $m_\uparrow^* \neq m_\downarrow^*$, for $\mathbf{Q}=0$ and $\mathbf{Q}\neq 0$, respectively. The dashed lines describe the same energy when $m_\uparrow^*=m_\downarrow^*=m_{av}$. **b,** Centre-of-mass momentum **Q** of the Cooper pair as a function of the field. **c,** and **d,** Binding energy *versus* $|\mathbf{Q}|$ for different values of the field. The tilted dotted lines represent the binding energy in the case when the total pair wave function is antisymmetric. **e,** Asymmetry factor of the spatial part of the pair wave function in the applied field describing the degree of deviation from its symmetric form with respect to $\mathbf{r}_1 \leftrightarrow \mathbf{r}_2$ transposition.

Fig. 3.

Mass splitting vs. H for square lattice for selected values of both the filling $n_f$ and temperature. The filling $n_f$ models an almost integer (3+) valency of $Ce^{(4-n_f)+}$ ion in the heavy fermion system CeCoIn$_5$. The effective masses depend on temperature via the system magnetization.

Fig. 4.

Phase boundaries for a superconductor with both the spin independent masses (**a**) and with the spin dependent mass (**b**), The FFLO phase is much more pronounced in the case with $m_\uparrow^* \neq m_\downarrow^*$, as the pair binding energy is larger even though we start from the same $m_\uparrow^* = m_{av}$ at H=0. The transition BCS→FFLO is discontinuous, whereas that between FFLO and the normal state is in the case (**a**) continuous and discontinuous in the case (**b**). **c,** and **d,** Gap magnitude and |**Q**| on the H-T plane, respectively, in the case corresponding to the phase diagram (**a**). The dashed lines represent the upper critical field ($H_{C2}$) for BCS state in the two situations.

Fig. 5.

Free-energy shape on Q plane for the BCS state and the FFLO state. The two separate minima lead to a discontinuous switching from BCS (**a**) to FFLO state (**b**). Note also a small difference in the free energy between the states; this may lead to a substantial blurring of the experimental detection of the phase boundary and mimic its quasicontinuous character with a small hysteresis.

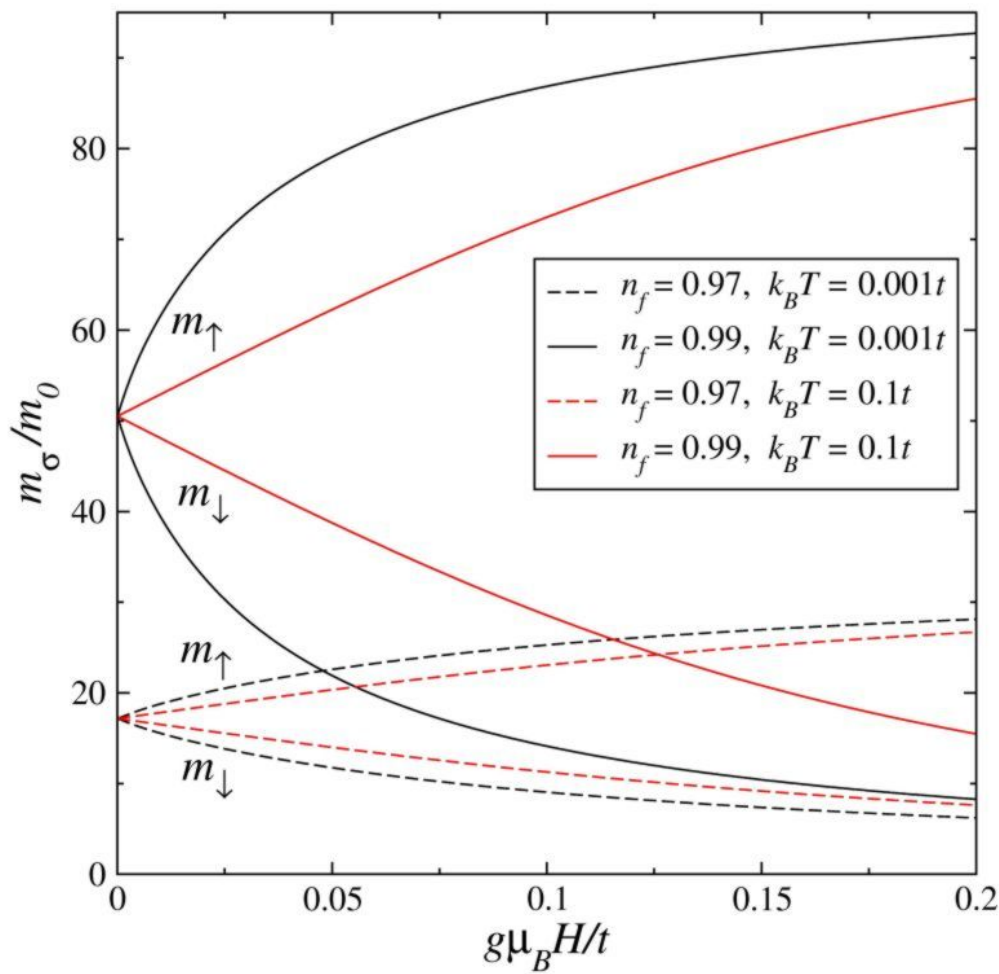

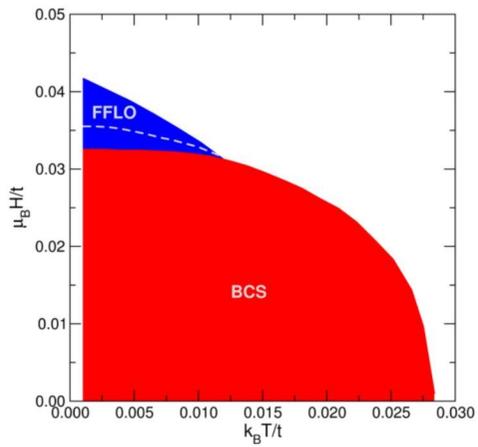
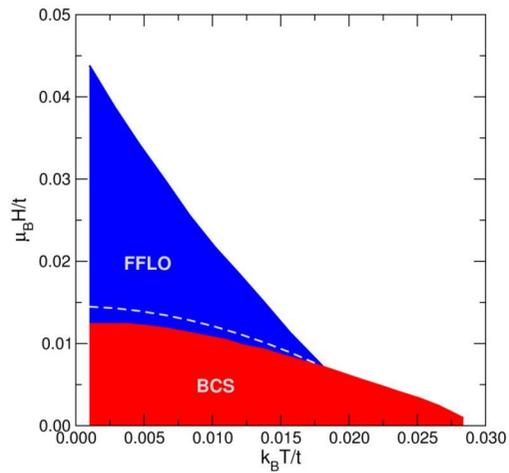
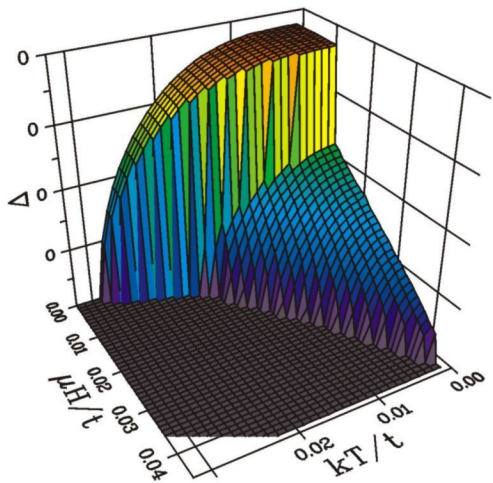
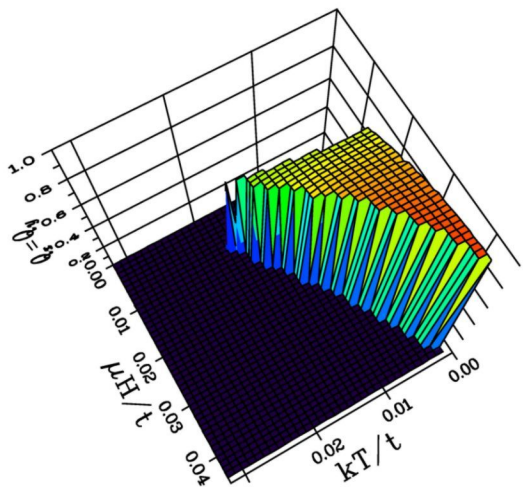

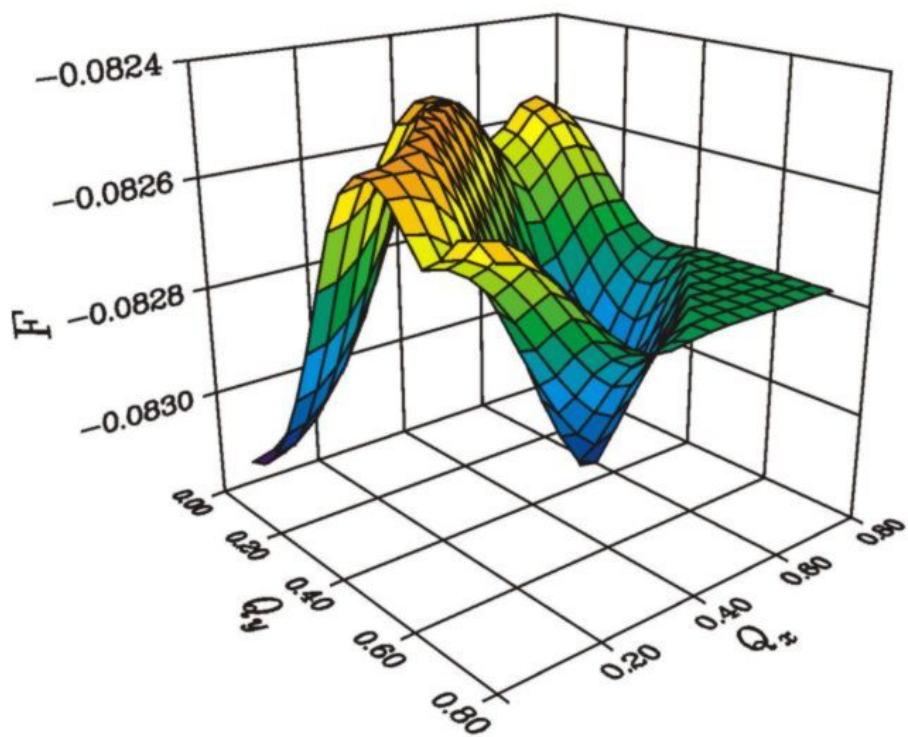
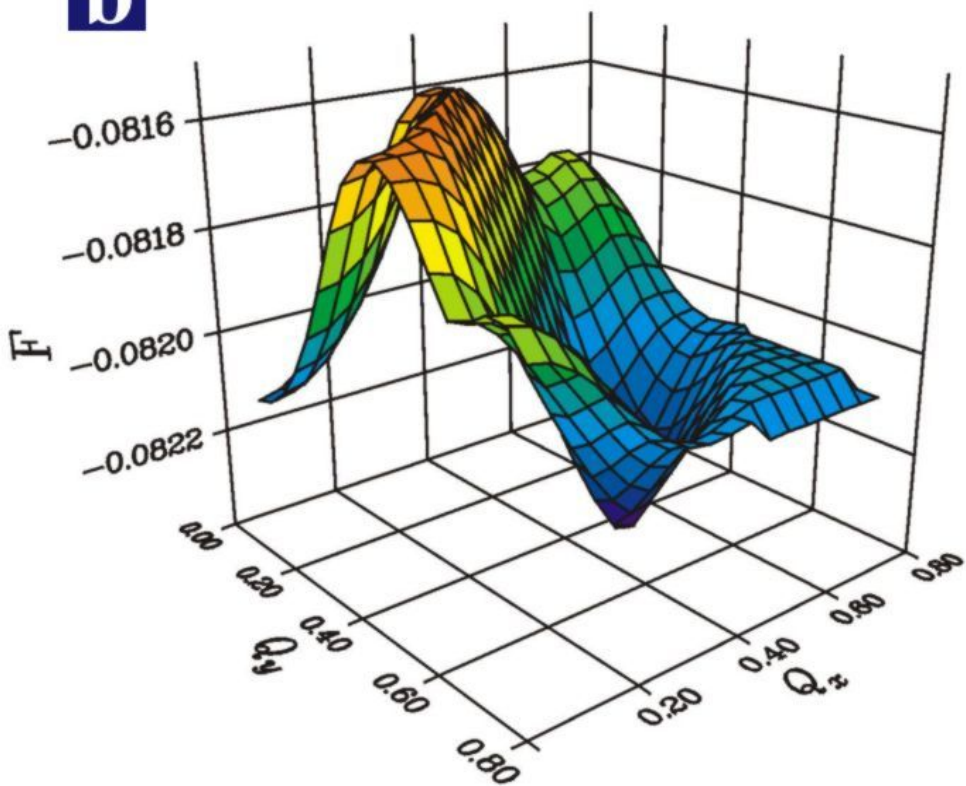

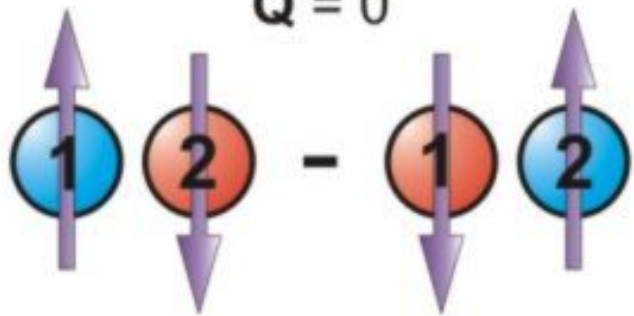 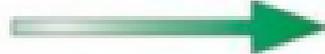 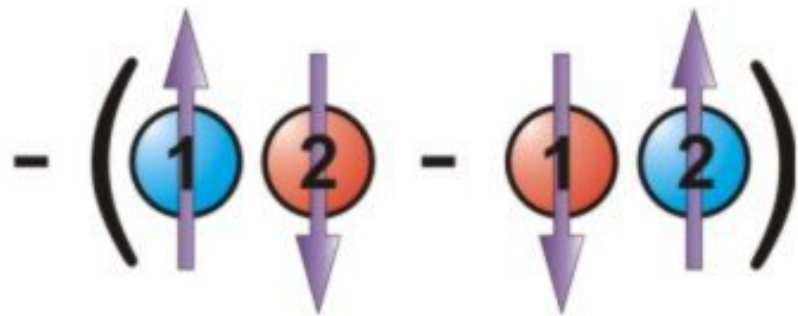
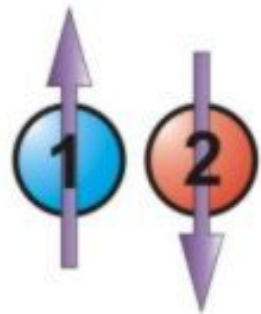 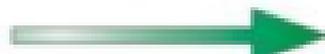 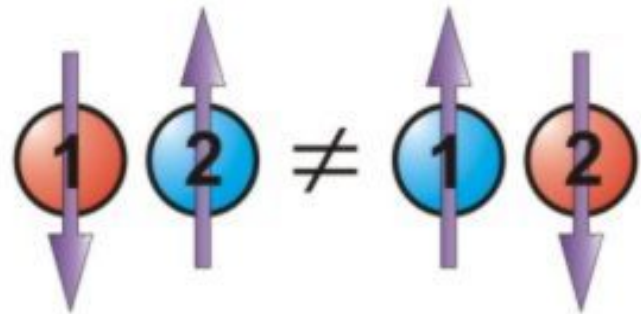

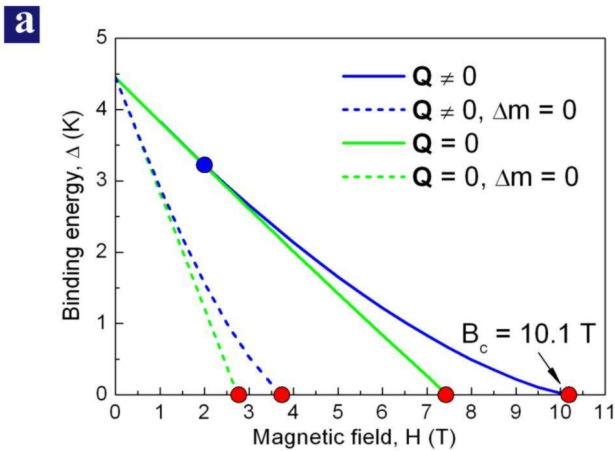
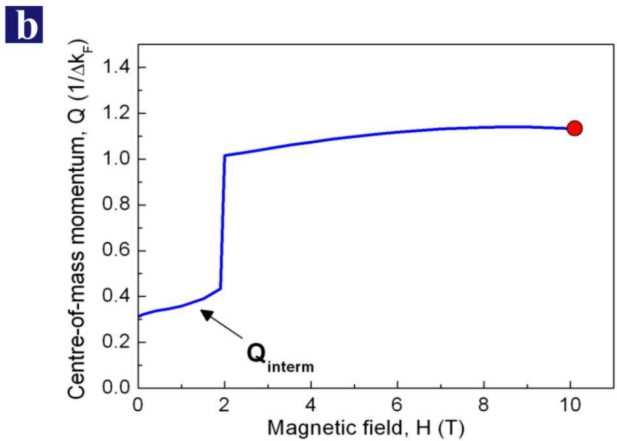
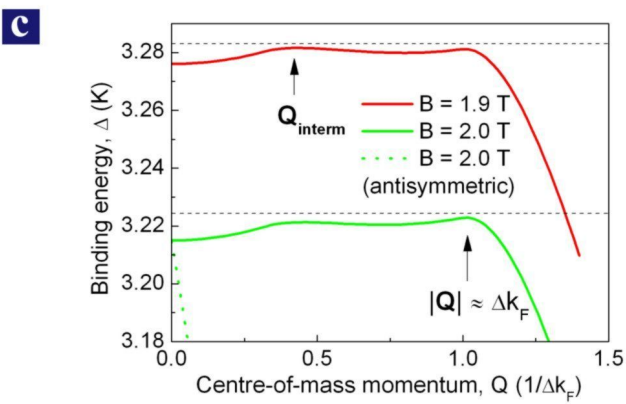
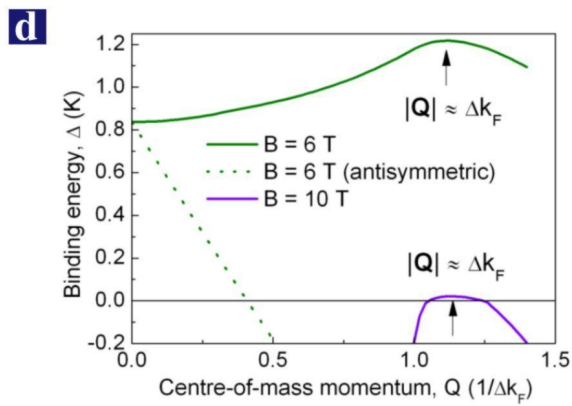
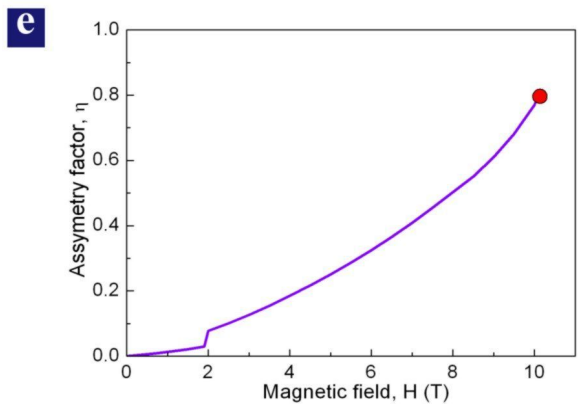